\documentclass{article} 
\usepackage{iclr2026_conference,times}

\usepackage{amsmath,amsfonts,bm}

\def\eqref#1{equation~\ref{#1}}

\def\1{\bm{1}}

\DeclareMathAlphabet{\mathsfit}{\encodingdefault}{\sfdefault}{m}{sl}
\SetMathAlphabet{\mathsfit}{bold}{\encodingdefault}{\sfdefault}{bx}{n}

\usepackage{hyperref}
\usepackage{url}
\usepackage{booktabs}       
\usepackage{multirow}       
\usepackage{amsfonts}       
\usepackage{nicefrac}       
\usepackage{microtype}      
\usepackage{xcolor}         
\usepackage{tikz}
\usetikzlibrary{shapes.geometric, arrows, positioning, fit, backgrounds, shadows}
\usetikzlibrary{decorations.pathreplacing}
\usetikzlibrary{shadows}
\usepackage{algorithm}
\usepackage{algorithmic}
\usepackage{xspace}
\usepackage{tcolorbox}
\tcbuselibrary{listings, breakable}
\definecolor{leankeyword}{rgb}{0.0, 0.0, 0.5} 
\definecolor{leancomment}{rgb}{0.0, 0.5, 0.0} 
\definecolor{leanstring}{rgb}{0.6, 0.1, 0.1}  

\lstdefinelanguage{lean}{
  morekeywords={def, theorem, lemma, structure, inductive, instance, where,
    calc, by, exact, have, obtain, let, if, then, else, match, with, namespace, section, end, open, set_option, variable, forall, exists, Prop, Type, Sort, Fintype, Nonempty},
  sensitive=true,
  morecomment=[l]{--},
  morecomment=[n]{/-}{-/},
  morestring=[b]",
  basicstyle=\ttfamily\footnotesize,
  keywordstyle=\bfseries\color{leankeyword},
  commentstyle=\itshape\color{leancomment},
  stringstyle=\color{leanstring},
  showstringspaces=false,
  breaklines=true,
  tabsize=2,
}
\newtcblisting{leancode}{
  colback=gray!10,
  colframe=gray!50,
  listing only,
  listing options={language=lean},
  left=5mm,
  top=2mm,
  bottom=2mm,
  breakable
}

\definecolor{teal}{RGB}{0,128,128}
\newcommand{\projname}{Quantum Topology\xspace}

\title{MerLean: An Agentic Framework For Autoformalization in Quantum Computation}

\author{%
  Yuanjie Ren\thanks{Equal contribution.} \\
  Massachusetts Institute of Technology\\
  Cambridge, MA 02139 \\
  \texttt{yuanjie@mit.edu} \\
  \And
  Jinzheng Li\footnotemark[1]\\
  Northeastern University\\
  Boston, MA 02115\\ 
  \texttt{li.jinzh@northeastern.edu} \\
  \AND
  Yidi Qi\footnotemark[1]\\
  Northeastern University \\
  Boston, MA 02115\\ 
  \texttt{y.qi@northeastern.edu} \\
}

\makeatletter
\def\section{\@startsection {section}{1}{\z@}{-2.0ex plus
    -0.5ex minus -.2ex}{1.5ex plus 0.3ex
minus0.2ex}{\large\scshape\raggedright}}
\def\subsection{\@startsection{subsection}{2}{\z@}{-1.8ex plus
-0.5ex minus -.2ex}{0.8ex plus .2ex}{\normalsize\scshape\raggedright}}
\def\subsubsection{\@startsection{subsubsection}{3}{\z@}{-1.5ex
plus      -0.5ex minus -.2ex}{0.5ex plus
.2ex}{\normalsize\scshape\raggedright}}
\makeatother

\iclrfinalcopy 
\begin{document}

\maketitle
\begin{center}
    \ificlrfinal
        \includegraphics[width=2.cm]{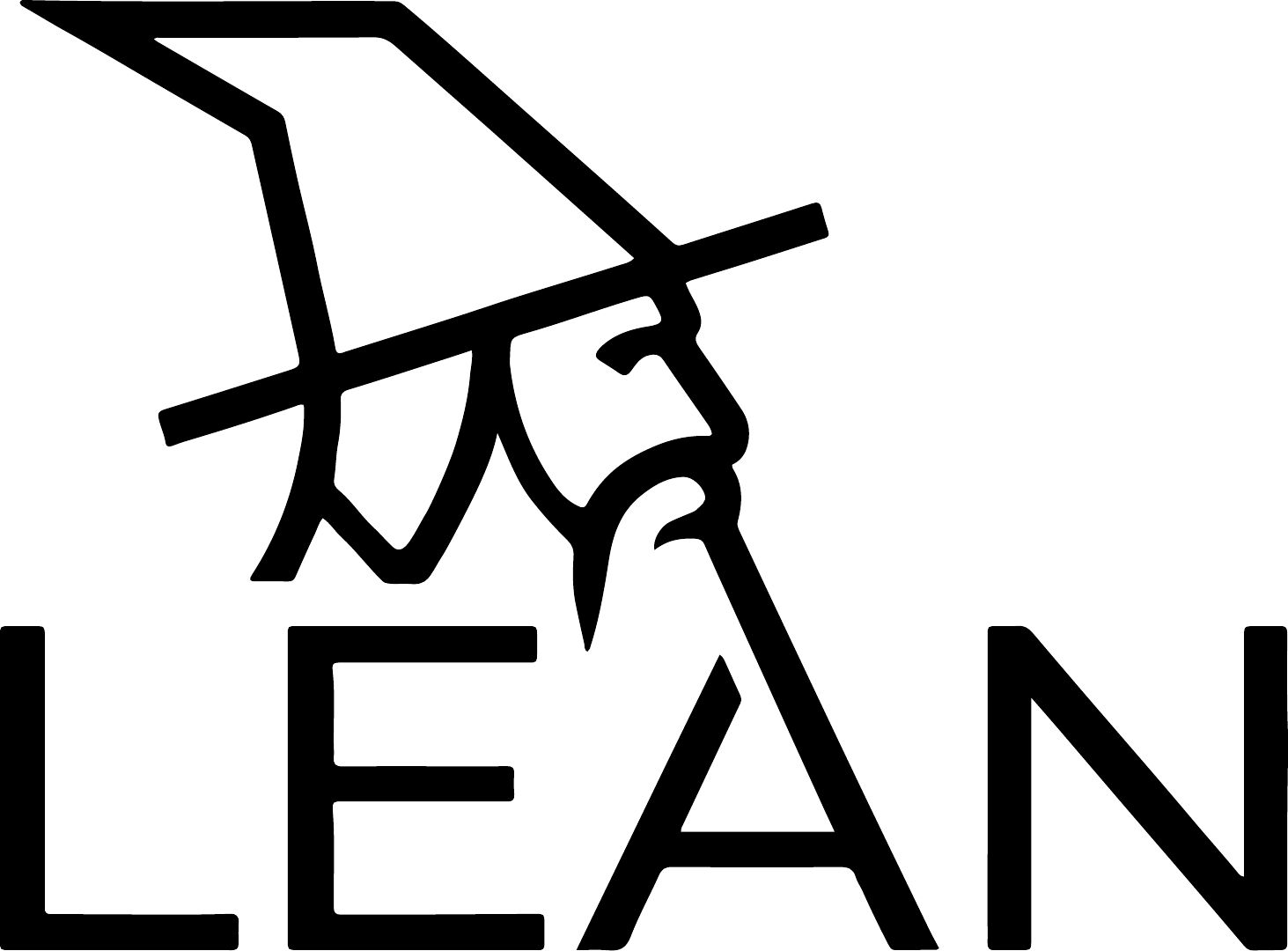}
    \fi
\end{center}
\begin{abstract}
We introduce MerLean, a fully automated agentic framework for autoformalization in quantum computation. MerLean extracts mathematical statements from \LaTeX{} source files, formalizes them into verified Lean~4 code built on Mathlib, and translates the result back into human-readable \LaTeX{} for semantic review. We evaluate MerLean on three theoretical quantum computing papers producing 2,050 Lean declarations from 114 statements in total. 
MerLean achieves end-to-end formalization on all three papers, reducing the verification burden to only the newly introduced definitions and axioms. Our results demonstrate that agentic autoformalization can scale to frontier research, offering both a practical tool for machine-verified peer review and a scalable engine for mining high-quality synthetic data to train future reasoning models. 
Our approach can also be generalized to any other rigorous research in mathematics and theoretical physics.
\end{abstract}

\section{Introduction}
The goal of autoformalization~\citep{szegedy_autoformalization} is to automatically translate mathematical statements from natural language into formal code verifiable by systems such as Isabelle~\citep{paulson_isabelle_1994}, Coq~\citep{team_coq_2024}, and Lean~\citep{moura_lean_2021}. Recent work has shown that Large Language Models (LLMs) are well suited to this task due to their semantic processing  capabilities~\citep{wu_autoformalization_2022,jiang_draft_2022, tarrach_more_2024, weng_autoformalization_2025}.

While most research in autoformalization focuses on pure mathematics, applications in physics have also attracted significant attention: Lean libraries such as PhysLean/HepLean~\citep{toobysmith2024hepleandigitalisinghighenergy} and Lean-QuantumInfo~\citep{meiburg2025formalization} are currently under development, with the latter specifically utilizing autoformalization tools to accelerate the process. Theoretical quantum computation serves as an ideal testbed for this workflow. The volume of submissions to the arXiv \texttt{quant-ph} category reached 11,891 articles in 2025~\citep{arxiv_stats}, creating a verification bottleneck that threatens to outpace the peer-review system. Functionally, the field operates as a subfield of applied mathematics, exhibiting deep connections to linear algebra, graph theory, and algebraic topology. The rigorous nature of these results makes them suitable for neuro-symbolic verification, while strong industrial interest ensures practical impact.

In this work, we introduce \textbf{MerLean}, a fully automated agentic framework designed to extract, organize, and formalize mathematical statements from \LaTeX{} source files, operating without human-in-the-loop intervention during the formalization process. After formalization succeeds, we run a second agent that ``autoinformalizes'' the verified code back into natural language, generating a blueprint for human experts to review the semantic alignment between the formal code and the original intent. We evaluate MerLean on three papers in theoretical quantum computing: one \textit{unpublished} manuscript (guaranteeing zero data contamination) and two previously published works. MerLean achieved end-to-end formalization on all three papers, introducing necessary definitions and axioms where results depend on mathematical machinery not yet available in Mathlib. 
\ificlrfinal
An interactive example of the formalization output is available at \url{https://doxtor6.github.io/MerLean-examples/}.
\fi

The scope of this framework extends beyond quantum computation. Any discipline that relies on formal mathematical proofs can use this neuro-symbolic approach to scale verification and accelerate discovery.

\section{Related Works}

LLM-based autoformalization and automated theorem provers demonstrated promising capabilities on competition-level problems~\citep{wu_autoformalization_2022, jiang_draft_2022}. Subsequent work improved upon these foundations by addressing proof detail expansion~\citep{tarrach_more_2024} and faithfulness verification~\citep{li_autoformalize_2024, liu_rethinking_2025, peng_criticlean_2025}. Extensions to domain-specific applications include Euclidean geometry~\citep{murphy_autoformalizing_2024}, biomedical text~\citep{zhang_autoformalization_2025}, and physics~\citep{zhang_physprover_2026}.

Recent work has shifted toward agentic systems that couple frontier LLMs with tool integration for interactive theorem proving. \citet{breen_ax-prover_2025} demonstrated Ax-Prover, an MCP-based multi-agent system using Claude Sonnet that outperforms specialized provers on out-of-domain benchmarks. \citet{xu_agentic_2026} demonstrated ``agentic proof automation'' by mechanizing System Capless (14,000+ lines of Lean~4) with 87\% task success rate across 189 annotated tasks; their workflow is \textit{human-guided}, with humans providing definitions, theorems, and proof strategies while agents handle mechanical proof engineering. \citet{liu2026numinaleanagentopengeneralagentic} proposed Numina-Lean-Agent, combining Claude Code with MCP-based Lean tools including Lean-LSP-MCP for goal querying and theorem search; it solved all 12 Putnam~2025 problems without model training and formalized the Brascamp--Lieb theorem (8,000+ lines), though again requiring substantial human guidance.

Architecturally, MerLean closely resembles both systems---sharing the paradigm of agentic interaction with Lean~4 in a generate-check-refine loop, frontier LLMs as the reasoning backbone, and the ambition of formalizing research-level mathematics. The primary feature is that MerLean achieves full-paper and fully automated end-to-end formalization without human guidance on the domain we tested.

\section{Framework}

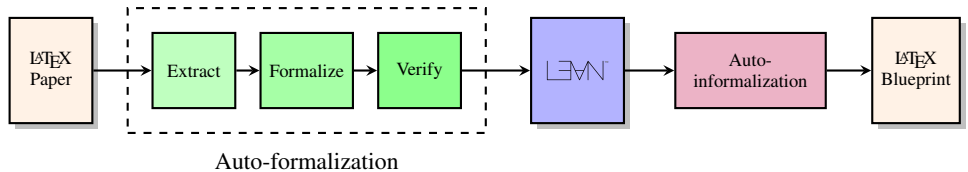
\begin{figure}[h]
    \centering
    \begin{tikzpicture}[
        layer/.style={rectangle, draw, thick, minimum height=1.0cm, minimum width=1.1cm, align=center, font=\scriptsize},
        doc/.style={rectangle, draw, thick, fill=white, drop shadow, minimum height=1.4cm, minimum width=1.1cm, align=center, font=\scriptsize},
        arrow/.style={->, >=stealth, thick},
        label/.style={font=\scriptsize}
    ]

    \node[doc, fill=orange!10] (in) at (-0.1,0) {\LaTeX{}\\Paper};

    \node[layer, fill=green!25] (e1) at (1.8,0) {Extract};
    \node[layer, fill=green!35] (e2) at (3.3,0) {Formalize};
    \node[layer, fill=green!45] (e3) at (4.8,0) {Verify};
    \node at (3.3,-1.2) {\small Auto-formalization};
    \node[doc, fill=blue!30] (z) at (6.9,0) {\includegraphics[width=1.0cm]{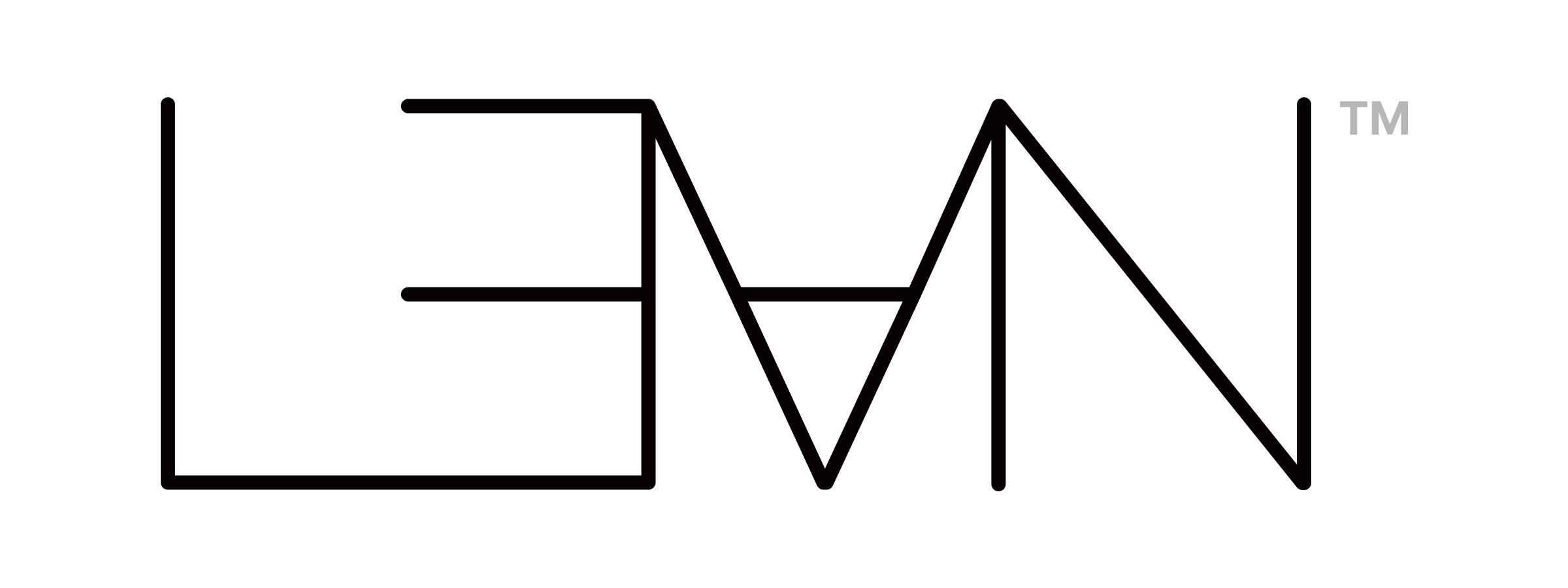}};

    \node[layer, minimum width=2.0cm, fill=purple!30] (dec) at (9.2,0) {Auto-\\informalization};

    \node[doc, fill=orange!10] (out) at (11.4,0) {\LaTeX{}\\Blueprint};

    \draw[arrow] (in) -- (e1);
    \draw[arrow] (e1) -- (e2);
    \draw[arrow] (e2) -- (e3);
    \draw[arrow] (e3) -- (z);
    \draw[arrow] (z) -- (dec);
    \draw[arrow] (dec) -- (out);

    \node[draw, dashed, thick, inner sep=9pt, fit=(e1) (e3), label=below:\small\textit{Auto-formalization}] {};

    \end{tikzpicture}
    \caption{MerLean architecture. \emph{Autoformalization} extracts statements from a \LaTeX{} paper, formalizes them into Lean~4, and verifies faithfulness. \emph{Autoinformalization} translates the verified code back into a human-readable \LaTeX{} blueprint or file.}
    \label{fig:architecture}
    \end{figure}

MerLean is a bidirectional autoformalization framework comprising two complementary pipelines: \emph{autoformalization} translates mathematical research papers from \LaTeX{} into verified Lean~4 libraries built on Mathlib, while \emph{autoinformalization} converts the formal code back into human-readable \LaTeX{} for expert review. This round-trip design ensures both machine-verified correctness and human-verifiable semantic alignment. The framework employs a frontier LLM agent based on Claude Code that engages multi-turn agentic interactions---iteratively refining outputs based on compiler feedback, faithfulness checks, and tool-augmented exploration of Mathlib. Figure~\ref{fig:architecture} illustrates the overall architecture.

\subsection{Autoformalization}

\paragraph{Statement Extraction.} Given a \LaTeX{} paper, the agent first extracts all mathematical \emph{statements}, including definitions, theorems, lemmas, propositions, corollaries, and remarks into a structured JSON representation. Each statement corresponds to a single mathematical result from the paper. It includes a unique identifier (e.g., \texttt{Def\_1}, \texttt{Thm\_2}), the mathematical content in natural language, explicit dependencies on other statements, and proof sketches when available. The extraction runs for multiple iterations, where each pass reviews and refines the previous output: expanding vague phrases like ``by standard arguments'' into concrete proof steps, adding missing intermediate lemmas, and ensuring statements are ordered so that dependencies precede dependents.

\paragraph{Iterative Formalization.} For each extracted statement, the agent enters a compile-fix loop. It generates Lean~4 code consisting of one or more \emph{declarations}, writes them to the appropriate file in the library structure, and compiles it. Each statement typically produces multiple declarations, as the agent autonomously introduces auxiliary definitions, helper lemmas, and typeclass instances to support the target result. If compilation fails, the error messages (type mismatches, unknown identifiers, tactic failures, etc.) will be parsed and fed back to the agent, which analyzes the errors and produces a corrected version. This loop continues until either the code compiles without errors and warnings, or until it reaches a maximum attempt limit. During this process, the agent has access to several diagnostic tools provided via \texttt{lean-lsp-mcp}, a Model Context Protocol server that exposes Lean's language server capabilities directly to the agent, \texttt{lean\_goal} for inspecting proof states at specific positions, \texttt{lean\_hover\_info} for type signatures and documentation, and semantic search tools (\texttt{leansearch}, \texttt{loogle}) for discovering relevant Mathlib lemmas. The agent can also grep through Mathlib source code to find usage patterns and verify that referenced lemmas actually exist.
\paragraph{Faithfulness Checking.} Compilation alone is insufficient: an LLM can produce code that type-checks but misrepresents the mathematics (e.g., proving a trivial statement instead of the intended theorem) \citep{lin2025goedelproverv2scalingformaltheorem}. 
After a successful build, MerLean 
reflects on  whether the result matches the original meaning. If yes, the program continues to the next statement; otherwise it continues to attempt.

\paragraph{Axiom Handling.} Frontier research often depends on results not yet in Mathlib (e.g., the K\"unneth formula for certain coefficient rings). MerLean handles this through explicit \texttt{axiom} declarations, clearly distinguishing intentional assumptions from incomplete proofs (\texttt{sorry}). When formalization fails after maximum attempts, an ``axiom phase'' converts blocking subgoals to axioms, producing partial formalizations with transparent assumptions that can be filled in as Mathlib expands.

\subsection{Autoinformalization}

The decoder reverses the formalization process, converting a verified Lean~4 library back into human-readable \LaTeX{}. While existing tools such as \texttt{doc-gen4} can produce reasonably readable documentation from Lean source code, the output remains deeply tied to Lean's type-theoretic syntax and is difficult to interpret for readers without formal methods expertise. Therefore, MerLean utilizes the same LLM model that performed the formalization to do the informalization: since the model has demonstrated its ability to formalize content in the specific research area, it should also be capable of translating each declaration into natural language, making it accessible to domain experts with no prior knowledge of Lean. Of course, no original statement or paper content is provided to the informalization agent to prevent data leaks. The pipeline parses all Lean files, constructs a dependency graph, and produces two complementary outputs: an interactive blueprint compatible with \texttt{leanblueprint} for web-based exploration of the dependency structure, and a standalone textbook-style narrative for readers unfamiliar with formal methods. Any unverified assumptions (\texttt{axiom} declarations) are prominently highlighted, ensuring full transparency about the boundaries of the formalization.

\section{Experiments}

We evaluate MerLean on three papers in theoretical quantum computing: one unpublished manuscript to guarantee zero data contamination, and two published papers to assess performance on existing literature.

\begin{itemize}
    \item \textbf{Paper A: Balanced Product Codes~\citep{breuckmann2021balanced}.} This paper studies quantum codes constructed from tensor products and fiber bundles of chain complexes. The mathematical machinery includes homological algebra, tensor product of complexes, expander graphs and spectral expansion.
    \item \textbf{Paper B: Fault-Tolerant Quantum Computation~\citep{williamson2024faulttolerant}.} This paper provides a comprehensive treatment of stabilizer codes and fault-tolerant protocols. The mathematical content includes stabilizer formalism and Pauli algebra, transversal gates, gauging graphs, fault-tolerant state preparation and measurement.
    \item \textbf{Paper C: \projname.} This is an unpublished manuscript, ensuring the content has never appeared in any LLM training data. The manuscript proves several algebraic and group-theoretic properties of some map on quantum computational systems. Details will be released later.
\end{itemize}

These three projects demonstrate the agent's capability to formalize and bridge logical gaps in frontier rigorous research, regardless of whether the content is present in the base LLM's training data. 
\ificlrfinal
An interactive example of the Fault-Tolerant QC formalization output is available at \url{https://doxtor6.github.io/MerLean-examples/}.
\fi

\subsection{Results}

\begin{table}[h]
\centering
\caption{Summary of formalization experiments per paper.}
\label{tab:summary}
\begin{tabular}{lcccc}
\toprule
\textbf{Paper} & \textbf{Statements} & \textbf{Lines of Lean} & \textbf{Declarations} & \textbf{Time}  \\
\midrule
Balanced Product  & 44 & 14,997 & 730 & 20h 4m  \\
Fault-Tolerant QC  & 47 & 18,557 & 923 & 11h 41m \\
\projname  & 23 & 7,761 & 397 & 7h 51m  \\
\bottomrule
\end{tabular}
\end{table}

\begin{table}[h]
\centering
\caption{Formalization statistics by statement type across all three papers.}
\label{tab:detailed_stats}
\begin{tabular}{lccc}
\toprule
\textbf{Type} & \textbf{Count} & \textbf{Avg. Time} & \textbf{Avg. Compiles}  \\
\midrule
Definition & 49 & 18m 0s & 11.7  \\
Theorem & 15 & 39m 41s & 22.4  \\
Lemma & 20 & 33m 22s & 18.3  \\
Remark & 26 & 10m 34s & 7.1  \\
Corollary & 4 & 19m 23s & 5.5  \\
\midrule
\textbf{Total/Avg} & 114 & 21m 54s & 13.0  \\
\bottomrule
\end{tabular}
\end{table}

Across all three papers, MerLean formalized 114 statements totaling 2,050 Lean declarations in under 42 hours of wall-clock time. 
Supported by manual review to ensure all new definitions and axioms are mathematically accurate and rigorously constructed, MerLean successfully formalized all three papers, demonstrating its capability on both novel and published content.
The Balanced Product Codes paper required explicit axioms for 9.1\% of statements, corresponding to results depending on machinery not yet in Mathlib (e.g., spectral sequences, K\"unneth isomorphisms for $\mathbb{F}_2$-chain complexes). Theorems were the hardest to formalize, averaging 39m 41s and 22.4 compile attempts, while remarks were the easiest at 10m 34s and 7.1 compiles with no axioms required. A representative fully-proved theorem is shown in Appendix~\ref{app:theorem}.

\begin{figure}[h]
    \centering
    \includegraphics[width=\textwidth]{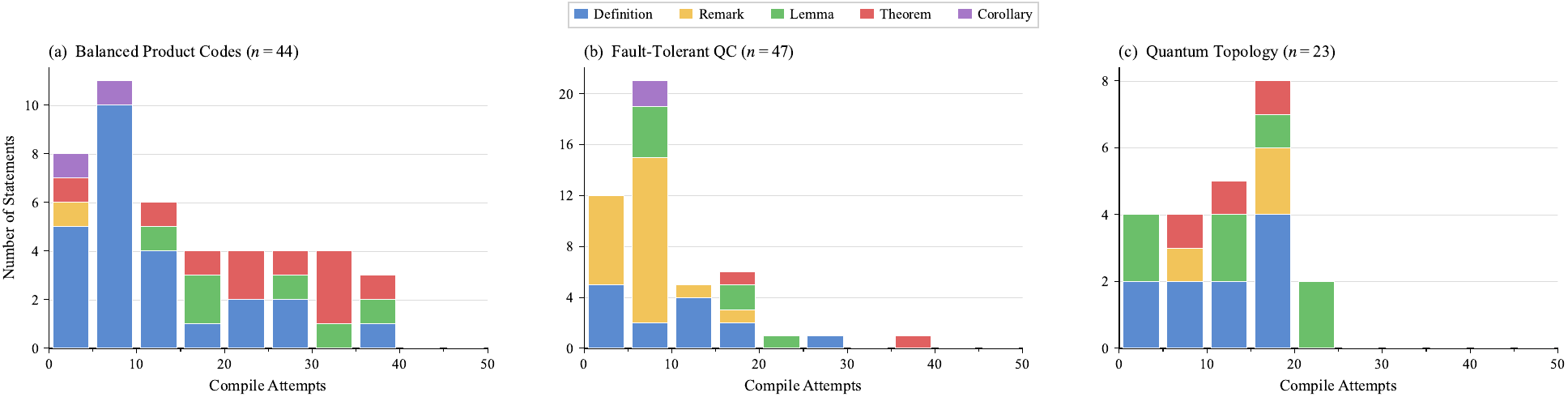}
    \caption{Distribution of compile attempts by statement type for each paper (maximum 30 attempts before axiom phase). Most statements resolve within 1--10 attempts; theorems and lemmas require significantly more iterations.}
    \label{fig:compile_hist}
\end{figure}
Figure~\ref{fig:compile_hist} shows the distribution of compile attempts per statement, broken down by type and paper. The distribution is heavily right-skewed across all three papers: most statements compile within 1--10 attempts, but a long tail of ``hard'' statements requires 21+ iterations. Balanced Product Codes exhibits the widest spread, reflecting its reliance on advanced algebraic machinery (spectral sequences, K\"unneth formulas) not yet in Mathlib. Fault-Tolerant QC is heavily concentrated in the low-compile regime, with 23 of 47 statements resolving within 5--10 attempts; its two theorems are the only statements exceeding 20 compiles. Quantum Topology shows a more uniform distribution, consistent with its smaller but technically diverse statement set. Across all papers, theorems and lemmas dominate the high-compile bins, while definitions and remarks cluster in the lower range.

\subsection{Qualitative Observations}

\paragraph{Compile-Fix Loop Behavior.} The hard statements that require 20+ compile attempts typically involve:
\begin{enumerate}
    \item Dependent type arithmetic (e.g., proving $i - 1 + 1 = i$ at the type level)
    \item Missing lemmas that require creative workarounds
    \item Tactic timeouts requiring proof restructuring
\end{enumerate}

\paragraph{Agent-Discovered Lemmas.} MerLean frequently introduces auxiliary lemmas not stated in the original paper. These helper lemmas are natural intermediate steps that a human mathematician would typically leave implicit (see Appendix~\ref{app:lemma} for a full example). For instance, the agent introduced:
\begin{itemize}
    \item \texttt{edgeBoundary\_card\_eq\_edgeCount} (\textit{Balanced Product Codes}): Bridges the gap between the geometric definition of expansion (boundary size) and algebraic adjacency counts, enabling the formal derivation of the Relative Cheeger Inequality.
    \item \texttt{pauliPair\_anticommuting\_ct\_satisfied} (\textit{Fault-Tolerant QC}): Algebraically verifies that anticommuting spacetime faults cancel out disjointly in the detector model, a necessary intermediate result for proving the Global Correctness of Spacetime Syndrome Extraction.
    \item \texttt{weight\_inequality\_core} (\textit{Fault-Tolerant QC}): Establishes that $w' - |S| + |\partial S| \geq \min(h(G),1)\cdot d$ by case-splitting on the Cheeger parameter $h(G)\geq 1$, bridging the cleaning lemma and the space distance lower bound (see Appendix~\ref{app:lemma}).
\end{itemize}

\paragraph{Axiom Declarations.} The Balanced Product Codes formalization required explicit axiom declarations where the underlying mathematical machinery is not yet available in Mathlib. Notably, the original paper invokes these classical theorems directly without proof, as they are well-known results in the literature; the informal proofs are therefore absent from the source material from the very beginning, making axiomatization the natural counterpart in the formal setting. These axioms reveal concrete gaps relevant to physics applications:
\begin{itemize}
    \item \textbf{K\"unneth Formula}: The isomorphism $H_n(C \otimes D) \cong \bigoplus_{p+q=n} H_p(C) \otimes H_q(D)$ for chain complexes over $\mathbb{F}_2$ is not available in Mathlib (\texttt{kunnethMap\_injective\_aux}, \texttt{kunnethMap\_surjective\_aux}).
    \item \textbf{Tensor-Homology Commutativity}: The isomorphism $H_q(V \otimes C) \cong V \otimes H_q(C)$ for a flat module $V$ is assumed via the combined axiom \texttt{e2PageIsoHomologyTensor\_ax}. This conceptually bundles the vertical and horizontal homology steps to identifying the spectral sequence $E^2$ pages with tensor products of base and fiber homologies in Theorem 3. The separate axioms \texttt{verticalHomologyIsoTensor\_ax} and \texttt{horizHomologyIsoTensor\_ax} are defined but superseded by this combined form in the main proof.
    \item \textbf{Spectral Sequences}: The machinery for computing homology of total complexes via spectral sequences is incomplete (\texttt{spectralSequenceIsomorphism\_nontrivial}). This is used in Theorem 3's proof of the Fiber Bundle K\"unneth Theorem to relate the $E^2$ page to the total homology.
\end{itemize}
These axioms are transparently marked in both the Lean code and the generated blueprint, so that readers immediately see which results rest on unverified assumptions. Appendix~\ref{app:axiom} shows the full K\"unneth formula formalization as a concrete example.

\section{Discussion}
\subsection{Challenges in MerLean}
\paragraph{Mathlib Gaps and Axiom-Based Formalization.} Our evaluation highlighted gaps in Mathlib regarding specialized physics concepts, for instance, the K\"unneth formula was unavailable. MerLean handles this by explicitly declaring \texttt{axiom} nodes in the dependency graph. More broadly, many physics concepts are not rigorously defined in the language of mathematics (e.g., notions in quantum field theory). Yet, it is out of the scope of this work (and similar ones) to build lean code of full prior knowledge.
Consequently, it is reasonable to develop libraries based on ``axioms''. Such choices of axioms should not be too elementary for the reason explained above, and not too advanced either, trivializing the derivation of the main results.

\paragraph{Faithfulness Checking.} A critical challenge is ensuring that formalized code accurately reflects the original text, rather than merely compiling without errors. MerLean's faithfulness checking pipeline is used to address this problem, while the autoinformalized blueprint exposes the logical structure for human review. Together, these mechanisms eliminate hallucinations as much as possible, making it immediately apparent if the agent has proven a trivial variation or fabricated a definition to satisfy the compiler.

\subsection{Future Applications of MerLean}
\paragraph{Research Assistant.} During our initial formalization of the unpublished \projname{} paper, one lemma persistently required an axiom. Examining the autoinformalized blueprint revealed that an ambiguous definition, where a constraint the author had implicitly assumed was lacking, caused the problem. After fixing the \LaTeX{} source, MerLean produced a complete formalization. 
This illustrates how the bidirectional pipeline can help researchers improve the rigor of their setups, definitions, and  proofs.

\paragraph{Formalized Peer Review.} Our validation on an unpublished manuscript confirms that MerLean operates without pre-training on the specific content. We envision a workflow where autoformalization occurs locally during drafting. Just as \LaTeX{} became the standard for mathematical typesetting, agentic frameworks could make formal repositories a standard companion to static PDFs, transforming peer review dynamics: reviewers could rely on machine-verified guarantees, focusing strictly on novelty and scientific significance.

\paragraph{Synthetic Data Flywheel.} 
 Autoformalization also plays a critical role in training specialized theorem provers, as the quality of large-scale synthetic data depends directly on accurate formalization~\citep{xin2024deepseekproveradvancingtheoremproving, kumarappan_leanagent_2025}.
MerLean facilitates a virtuous cycle for LLM training by mining high-quality \textit{(natural language, formal code)} pairs grounded in the scientific research they are based on. Feeding it back into base models will improve the next generation of agents.

\paragraph{Contributing to Physics Libraries.} The Lean ecosystem has seen growing efforts to formalize physics, including PhysLean/HepLean~\citep{toobysmith2024hepleandigitalisinghighenergy} for general physics and Lean-QuantumInfo~\citep{meiburg2025formalization} for quantum information theory. MerLean can accelerate contributions to these libraries by formalizing relevant research papers and extracting reusable definitions and lemmas. Alternatively, researchers can use MerLean to create domain-specific libraries by formalizing a collection of strongly related papers, building a coherent formal foundation for their research area that captures the interconnected definitions, lemmas, and theorems spanning multiple works.

\section{Conclusion}

We have presented MerLean, a fully automated bidirectional framework for formalizing mathematical research papers from \LaTeX{} into verified Lean~4 libraries. Our evaluation on three papers in theoretical quantum computation, covering stabilizer codes, fault-tolerant protocols, balanced product codes, and homological algebra, produced over 2,000 Lean declarations across 41,000+ lines of verified code, demonstrating that fully automated formalization of frontier research is feasible. The iterative compile-fix-verify loop effectively produces faithful formalizations without human intervention, while the autoinformalization pipeline enables domain experts to review semantic alignment without formal methods expertise.

Looking ahead, while theoretical quantum computing serves as an effective testbed, its mathematical substrate, being primarily linear algebra and functional analysis, benefits from mature Mathlib support. We plan to extend our evaluation to other scientific domains and branches of pure mathematics characterized by deeper dependency chains or different foundational structures, such as algebraic geometry and number theory, to fully establish the framework's generalizability.

We also plan to evaluate MerLean on existing autoformalization benchmarks to enable direct comparison with prior work. However, current benchmarks (e.g., miniF2F, ProofNet) focus primarily on isolated theorem statements rather than full paper formalization with interconnected definitions, lemmas, and theorems. In the future, we intend to create a dedicated benchmark comprising diverse research papers across multiple mathematical domains, enabling systematic evaluation of autoformalization systems at the level of frontier research.

\section*{LLM Usage Disclosure}

In accordance with ICLR 2026 policy, we disclose the following uses of large language models in this work:
\paragraph{Research.} MerLean uses Claude (Opus 4.5) as its core reasoning engine for both autoformalization and autoinformalization. The LLM performs statement extraction from \LaTeX{}, Lean code generation, error diagnosis and repair, and natural language translation of formal code. All experimental results reported in this paper were produced by this LLM-based system. The authors have verified and validated all research contributions.
\paragraph{Writing.} Claude Code was used as a writing assistant for editing portions of this manuscript, including fixing typos and improving grammar. All content was reviewed, verified, and revised by the human authors, who take full responsibility for the accuracy and integrity of the final submission.

\bibliography{iclr2026_conference}
\bibliographystyle{iclr2026_conference}

\appendix

\section{Formalization Examples}

We provide representative examples from the Fault-Tolerant QC formalization to illustrate the quality of MerLean's output.

\subsection{Theorem: Fault Tolerance of Gauging Measurement}
\label{app:theorem}

Theorem~2 of \cite{williamson2024faulttolerant} (adapted): The fault-tolerant implementation of Algorithm~1 with a suitable graph $G$ has spacetime fault-distance $d_{\mathrm{ST}} = d$, where $d$ is the distance of the original code. Specifically, if (1) the Cheeger constant satisfies $h(G) \geq 1$, and (2) the number of measurement rounds satisfies $t_o - t_i \geq d$, then any spacetime logical fault has weight at least $d$.

\begin{leancode}
/-- Main Theorem (Fault Tolerance): The spacetime fault-distance
    equals d. Under the conditions:
    1. h(G) >= 1 (Cheeger constant at least 1)
    2. (t_o - t_i) >= d (sufficient measurement rounds)
    The spacetime fault-distance d_ST equals exactly d. -/
theorem FaultToleranceTheorem
    [Fintype V] [Fintype E] [Fintype M] [Nonempty M]
    (DC : DetectorCollection V E M)
    (baseOutcomes : OutcomeAssignment M)
    (logicalEffect : SpacetimeFault V E M -> Prop)
    (cfg : FaultToleranceConfig)
    (h_all_decompose : forall F,
      IsSpacetimeLogicalFault DC baseOutcomes logicalEffect F ->
      LowerBoundCase DC baseOutcomes logicalEffect cfg F)
    (h_exists_d : exists F,
      IsSpacetimeLogicalFault DC baseOutcomes logicalEffect F
      /\ F.weight (intervalRounds cfg) = cfg.d) :
    spacetimeFaultDistance DC baseOutcomes logicalEffect
      (intervalRounds cfg) = cfg.d := by
  -- Get the weight-d logical fault for existence
  obtain <F_d, hF_d_log, hF_d_weight> := h_exists_d
  -- Upper bound: d_ST <= d
  have h_le : spacetimeFaultDistance DC baseOutcomes
      logicalEffect (intervalRounds cfg) <= cfg.d := by
    calc spacetimeFaultDistance DC baseOutcomes
          logicalEffect (intervalRounds cfg)
        <= F_d.weight (intervalRounds cfg) :=
          spacetimeFaultDistance_le_weight DC baseOutcomes
            logicalEffect (intervalRounds cfg) F_d hF_d_log
      _ = cfg.d := hF_d_weight
  -- Lower bound: d_ST >= d
  have h_ge : spacetimeFaultDistance DC baseOutcomes
      logicalEffect (intervalRounds cfg) >= cfg.d := by
    -- Get the minimum-achieving fault
    have h_has : hasLogicalFault DC baseOutcomes
      logicalEffect := <F_d, hF_d_log>
    obtain <F_min, hF_min_log, hF_min_weight> :=
      spacetimeFaultDistance_is_min DC baseOutcomes
        logicalEffect (intervalRounds cfg) h_has
    -- Apply lower bound to F_min
    have h_min_ge := spacetimeFaultDistance_ge_d DC
      baseOutcomes logicalEffect cfg F_min hF_min_log
      (h_all_decompose F_min hF_min_log)
    calc cfg.d
        <= F_min.weight (intervalRounds cfg) := h_min_ge
      _ = spacetimeFaultDistance DC baseOutcomes
          logicalEffect (intervalRounds cfg) := hF_min_weight
  -- Combine
  omega
\end{leancode}

\subsection{Agent-Discovered Helper Lemma}
\label{app:lemma}

This lemma is part of the formalization of \cite{williamson2024faulttolerant}, specifically addressing the lower bound on space distance. The agent introduced this intermediate lemma on its own to complete the proof of the space distance lower bound in the fault-tolerant quantum computation formalization.  Given a Cheeger-like expansion parameter $h(G)$, a cleaning set of size at most the cleaned operator weight, and a boundary satisfying the isoperimetric inequality, the lemma establishes:
\[
  w' - |S| + |\partial S| \;\geq\; \min(h(G),\,1)\cdot d,
\]
where $w'$ is the cleaned weight, $|S|$ the cleaning set size, $|\partial S|$ the boundary size, and $d$ the code distance.  The proof splits on whether $h(G)\geq 1$, using a \texttt{calc} chain with \texttt{nlinarith} in each branch.

\begin{leancode}
theorem weight_inequality_core
    (hG : R) (hG_nonneg : 0 <= hG)
    (d : N) (_hd_pos : 0 < d)
    (cleanedWeight : N)
    (hCleaned : cleanedWeight >= d)
    (cleaningSetSize : N)
    (hCleaningBound : cleaningSetSize <= cleanedWeight)
    (boundarySize : N)
    (hCheeger : (boundarySize : R) >= hG * cleaningSetSize) :
    (cleanedWeight : R) - cleaningSetSize + boundarySize
      >= minCheegerOne hG * d := by
  simp only [minCheegerOne]
  by_cases hG_ge_1 : hG >= 1
  . -- Case h(G) >= 1: boundarySize >= cleaningSetSize, so result >= cleanedWeight >= d
    rw [min_eq_right hG_ge_1]
    have hBound : (boundarySize : R) >= cleaningSetSize := by
      calc (boundarySize : R) >= hG * cleaningSetSize := hCheeger
        _ >= 1 * cleaningSetSize := by nlinarith
        _ = cleaningSetSize := one_mul _
    have hCleaned' : (cleanedWeight : R) >= d :=
      Nat.cast_le.mpr hCleaned
    linarith
  . -- Case h(G) < 1
    push_neg at hG_ge_1
    rw [min_eq_left (le_of_lt hG_ge_1)]
    have hClean : (cleanedWeight : R) >= d :=
      Nat.cast_le.mpr hCleaned
    have hS' : (cleaningSetSize : R) <= cleanedWeight :=
      Nat.cast_le.mpr hCleaningBound
    calc (cleanedWeight : R) - cleaningSetSize + boundarySize
        >= cleanedWeight - cleaningSetSize
           + hG * cleaningSetSize := by linarith
      _ = cleanedWeight
          + (hG - 1) * cleaningSetSize := by ring
      _ >= cleanedWeight
          + (hG - 1) * cleanedWeight := by nlinarith
      _ = hG * cleanedWeight := by ring
      _ >= hG * d := by nlinarith
\end{leancode}

\subsection{Axiom Example: K\"unneth Formula}
\label{app:axiom}
This example is drawn from the formalization of \cite{breuckmann2021balanced}, where the K\"unneth formula is essential for calculating the homology of balanced product codes. The K\"unneth formula is a fundamental result in homological algebra that relates the homology of a tensor product of two chain complexes to the tensor products of their individual homologies.  Given chain complexes $C=(C_\bullet,\partial^C)$ and $D=(D_\bullet,\partial^D)$ over~$\mathbb{F}_2$, the formula states:
\[
  H_n(C \otimes D) \;\cong\; \bigoplus_{p+q=n} H_p(C) \otimes H_q(D).
\]
Over a field such as~$\mathbb{F}_2$, all modules are flat, so the Tor correction terms vanish and the K\"unneth map is an isomorphism.  The isomorphism sends a pair of homology classes $[z]\in H_p(C)$ and $[w]\in H_q(D)$ to the class $[z\otimes w]\in H_{p+q}(C\otimes D)$.  Proving this in Lean~4 requires (i)~showing the tensor product of cycles is a cycle, (ii)~showing boundaries are preserved so the map descends to homology, and (iii)~establishing bijectivity.  Steps (i)--(iii) involve unfolding the internal colimit structure of Mathlib's total complex, which is not yet supported; these are axiomatized.  The full formalization is shown below.

The 4 snippets below correspond to the definition of types, followed by the three proof steps (i)--(iii).

First, we define the source type (direct sum of tensor products of homologies) and the target type (homology of the tensor product):
\begin{leancode}
variable (C D : F2ChainComplex)

/-- All F2-modules are flat (free modules are flat). -/
instance flat_F2_module (M : Type*) [AddCommGroup M]
    [Module F2 M] : Module.Flat F2 M :=
  Module.Flat.of_free

/-- Tensor product of homology spaces. -/
noncomputable def HomologyTensor (p q : Z) : Type :=
  (C.Homology p) (x)[F2] (D.Homology q)

...

/-- Index set: pairs (p,q) with p + q = n. -/
def KunnethIndex (n : Z) : Type :=
  { pq : Z x Z // pq.1 + pq.2 = n }

/-- The direct sum (+)_{p+q=n} H_p(C) (x) H_q(D). -/
noncomputable def KunnethDirectSum (n : Z) : Type :=
  Pi_0 (i : KunnethIndex n),
    HomologyTensor C D i.val.1 i.val.2

...

/-- Homology of the tensor product complex at degree n. -/
noncomputable abbrev TensorHomology (n : Z) : Type :=
  (TensorProductComplex C D).Homology n
\end{leancode}

\emph{Step (i): Constructing the map on cycles.} We show that the tensor product of cycles is a cycle, allowing us to define the cross product on cycles:
\begin{leancode}
/-- Map from C_p (x) D_q to (C (x) D)_{p+q}. -/
noncomputable def tensorInclusion (p q : Z) :
    (C.X p) (x)[F2] (D.X q) ->l[F2]
      (TensorProductComplex C D).X (p + q) :=
  (TensorProductComplex.i C D p q).hom

/-- Axiom: z (x) w is a cycle when both z, w are cycles.
d(z (x) w) = dz (x) w +- z (x) dw = 0. -/
axiom cycle_tensor_cycle_is_cycle'_aux
    (C D : F2ChainComplex) (p q : Z)
    (z : C.X p) (hz : z in C.Cycles p)
    (w : D.X q) (hw : w in D.Cycles q) :
    tensorInclusion C D p q (z (xt) w) in
      (TensorProductComplex C D).Cycles (p + q)

/-- Map from C.Cycles p (x) D.Cycles q to total cycles. -/
noncomputable def cyclesCrossProduct (p q : Z) :
    (C.Cycles p) (x)[F2] (D.Cycles q) ->l[F2]
      (TensorProductComplex C D).Cycles (p + q) := by
  ...  -- bilinear map lifting cycle_tensor_cycle_is_cycle'
\end{leancode}
\emph{Step (ii): Descent to homology.} We axiomatize that boundaries are preserved under the cross product, ensuring the map descends to a well-defined map on homology:
\begin{leancode}
/-- Axiom: boundary (x) cycle is a boundary.
If z = dz', then z (x) w = d(z' (x) w) since dw = 0. -/
axiom boundary_tensor_cycle_is_boundary'_aux
    (C D : F2ChainComplex) (p q : Z)
    (z : C.Cycles p) (hz : z.val in C.Boundaries p)
    (w : D.Cycles q) :
    (cyclesCrossProduct C D p q (z (xt) w)).val in
      (TensorProductComplex C D).Boundaries (p + q)

/-- Axiom: cycle (x) boundary is a boundary.
If w = dw', then z (x) w = d(z (x) w') since dz = 0. -/
axiom cycle_tensor_boundary_is_boundary'_aux
    (C D : F2ChainComplex) (p q : Z)
    (z : C.Cycles p)
    (w : D.Cycles q) (hw : w.val in D.Boundaries q) :
    (cyclesCrossProduct C D p q (z (xt) w)).val in
      (TensorProductComplex C D).Boundaries (p + q)
\end{leancode}

\emph{Step (iii): Bijectivity and the Main Theorem.} Finally, we construct the K\"unneth map and axiomatize its bijectivity to establish the isomorphism:
\begin{leancode}
...

/-- Cross product descends to homology: [z] (x) [w] |-> [z (x) w]. -/
noncomputable def kunnethComponentMap (p q : Z) :
    HomologyTensor C D p q ->l[F2]
      TensorHomology C D (p + q) :=
  TensorProduct.lift (kunnethComponentMapAux2 C D p q)

/-- The Kunneth map from (+)_{p+q=n} H_p(C) (x) H_q(D)
to H_n(C (x) D). -/
noncomputable def kunnethMap (n : Z) :
    KunnethDirectSum C D n ->l[F2]
      TensorHomology C D n := by
  refine DFinsupp.lsum N ?_
  intro <<p, q>, hpq>
  subst hpq
  exact kunnethComponentMap C D p q

/-- Axiom: the Kunneth map is injective over F2. -/
axiom kunnethMap_injective_aux
    (C D : F2ChainComplex) (n : Z) :
    Function.Injective (kunnethMap C D n)

/-- Axiom: the Kunneth map is surjective over F2. -/
axiom kunnethMap_surjective_aux
    (C D : F2ChainComplex) (n : Z) :
    Function.Surjective (kunnethMap C D n)

/-- Kunneth Formula:
(+)_{p+q=n} H_p(C) (x) H_q(D) ~= H_n(C (x) D). -/
noncomputable def kunnethEquiv (n : Z) :
    KunnethDirectSum C D n ~=l[F2]
      TensorHomology C D n :=
  LinearEquiv.ofBijective (kunnethMap C D n)
    <kunnethMap_injective C D n,
     kunnethMap_surjective C D n>
\end{leancode}

\end{document}